\documentstyle[11pt]{article}
\input{psfig}
\long\def\comment#1{}
\begin{document}
\title{On \={A}ryabha\d{t}a's Planetary Constants}

\author{Subhash Kak\thanks{
Department of Electrical \& Computer Engineering,
Louisiana State University,
Baton Rouge, LA 70803-5901, USA,
Email: {\tt kak@ece.lsu.edu}}}

\maketitle

\begin{abstract}
This paper examines
the theory of a Babylonian 
origin of \={A}ryabha\d{t}a's planetary 
constants.
It shows that \={A}ryabha\d{t}a's basic constant
is closer to the Indian counterpart than to the
Babylonian one.
Sketching connections between \={A}ryabha\d{t}a's framework
and earlier Indic astronomical ideas on
yugas and cyclic calendar systems, it is argued that 
\={A}ryabha\d{t}a's system is an outgrowth of an
earlier Indic tradition. 

\end{abstract}

\section{Introduction}
An old problem in the history of Indian science is whether
ideas at the basis of \={A}ryabha\d{t}a's astronomy
were borrowed from outside or were part of India's
own tradition.
This problem was first raised in the context of the 
now discredited thesis that
sound observational astronomy did not exist in India
prior to India's encounter with the West.
Thus in a recent paper,$^1$ Abhyankar argues that
``\={A}ryabha\d{t}a's values of {\it bhaga\d{n}as}
were probably derived from the Babylonian planetary
data.'' 
But Abhyankar makes contradictory assertions in the
paper, suggesting at one place that \={A}ryabha\d{t}a
had his own observations and at another place
that he copied numbers without understanding, making
a huge mistake in the process.

In support of his theory, Abhyankar
claims that \={A}ryabha\d{t}a used
the Babylonian value of 44528
synodic months in 3600 years as
his starting point. But this value
is already a part of the \'{S}atapatha
altar astronomy reconciling lunar and
solar years in a 95-year yuga.
In this ritual, an altar is built to an area
that is taken to represent the nak\d{s}atra
or the lunar year in tithis and the
next design is the same shape but to a larger
area
(solar year in tithis), but since this second 
design is too large, the altar construction
continues in a sequence of 95 years.
It appears that satisfactory reconciliation
by adding intercalary months to the
lunar year of 360 tithis amounted to subtracting
a certain number of tithis from the
372 tithis of the solar year, whose most likely
value was 89 tithis in 95 years.$^2$

The areas of the altars increase from $7\frac{1}{2}$
to $101\frac{1}{2}$ in the 95 long sequence in 
increments of one. The average size of the altar
is therefore 
$54\frac{1}{2}$, implying that the average difference
between the lunar and the solar year is taken to
be one unit with 
$54\frac{1}{2}$ which is about $6.60$ tithis for
the lunar year of 360 tithis.
This is approximately correct.

Considering a correction of 89 tithis in 95
years,
the corrected length of the
year is $372 - 89/95 = 371.06316$ tithis.
Since each lunation occurs in 30 tithis, the number
of lunations in 3600 years is 44527.579.
In a Mah\={a}yuga, this amounts to
53,433,095. In fact, the number chosen by
\={A}ryabha\d{t}a (row 1 in Table 1)
is closer to this number
rather than the Babylonian number
of 53,433,600.

Table 1 presents the Babylonian numbers
given by Abhyankar together with
the \={A}ryabha\d{t}a constants 
related to the synodic lunar months and the
revolutions of the lunar node, the lunar apogee,
and that of the planets.
The so-called Babylonian numbers are not actually
from any Babylonian text but were computed by
Abhyankar using the rule of three on various
Babylonian constants.

\vspace{0.2in}
{\it Table 1:} Revolutions in one Mah\={a}yuga

\begin{tabular}{||r|r|r||} \hline
Type & Babylonian & \={A}rybha\d{t}a \\ \hline
Synodic lunar months & 53,433,600 & 52,433,336 \\ \hline
Lunar node & -232,616 & -232,352 \\ \hline
Lunar apogee & 486,216 & 488,219 \\ \hline
Mercury & 17,937,000 & 17,937,020 \\ \hline
Venus & 7,022,344 & 7,022,388 \\ \hline
Mars & 2,296,900 & 2,296,824 \\ \hline
Jupiter & 364,216 & 364,224 \\ \hline
Saturn & 146,716 & 146,564 \\ \hline
\end{tabular}

\vspace{0.2in}
We see that no numbers match.
How does one then make the case that
\={A}ryabha\d{t}a obtained
his numbers from a Babylonian 
text?
Abhyankar says that 
these numbers are
different because of his 
(\={A}ryabha\d{t}a's) 
own
observations
``which are more accurate.''
But if \={A}ryabha\d{t}a had his
own observations, why did he have to
``copy" Babylonian constants, and end up not using them,
anyway?

Certain numbers have great discrepancy,
such as those of the lunar apogee, which
Abhyankar suggests was due to a ``wrong
reading of 6 by 8'' implying--in opposition
to his earlier view in the same paper that \={A}ryabha\d{t}a
also had his own observations--
that \={A}ryabha\d{t}a did not
possess his own data and that he
simply copied numbers from some manual
brought from Babylon!

The \={A}ryabha\d{t}a numbers are also more
accurate that Western numbers as in the work
of Ptolemy.$^3$
Given all this, 
there is no credible case to accept
the theory of borrowing of these
numbers from Babylon.

Abhyankar further suggests 
that \={A}ryabha\d{t}a may
have borrowed from Babylon
the two central
features of his system: (i) the concept of
the Mah\={a}yuga, and (ii) mean 
superconjunction of all planets at
some remote epoch in time.
In fact, Abhyankar repeats here 
an old theory of Pingree$^4$ and
van der Waerden$^5$ about a
transmission from Babylon of these
two central ideas.
In this paper, we show that these ideas were already
present in the pre-Siddh\={a}ntic
astronomy and, therefore, a
{\it contrived} connection with
Babylonian tables is unnecessary.

\section{The Indic tradition of yugas and superconjunctions}

In the altar ritual 
of the Br\={a}hma\d{n}as,$^6$ 
equivalences by number connected the altar area to the
length of the year.
The 5-year yuga is described in the
Ved\={a}\.{n}ga Jyoti\d{s}a, where
only the motions of the sun and the moon
are considered.
The \'{S}atapatha Br\={a}hma\d{n}a 
describes the 95-year cycle to harmonize
the solar and the lunar years.
The \'{S}atapatha Br\={a}hma\d{n}a
also describes an asymmetric circuit for
the sun$^7$, which the Greeks speak about
only around 400 BC.

Specifically, we find mention of the nominal year of
372 tithis, the nak\d{s}atra year of 324 tithis, and a solar
year of 371 tithis.
The fact that a further correction was required in 95 years
indicates that these figures were in themselves considered to be
approximate.

In the altar ritual, the primal person is made to an area of
$7 \frac{1}{2}$ puru\d{s}as, when a puru\d{s}a is also equated with
360 years leading to another cycle of 2700 years.
This is the 
Saptar\d{s}i cycle which was taken to start and end with a 
superconjunction.

The \'{S}atapatha Br\={a}hma\d{n}a 10.4.2.23-24 describes
that the \d{R}gveda has 432,000 syllables,
the Yajurveda has 288,000 and the
S\={a}maveda has 144,000 syllables.
This indicates that larger yugas in proportion
of 3:2:1 were known at the time of the
conceptualization of the Sa\d{m}hit\={a}s.

Since the nominal size of the \d{R}gveda was considered to be
432,000 syllables (SB 10.4.2.23)
we are led to the theory of a much larger
yuga of that extent in years since the \d{R}gveda
represented the universe symbolically.

Elsewhere, I show$^8$ how the 
Ved\={a}\.{n}ga Jyoti\d{s}a serves as a coordinate
system for the sun and the moon in terms of
the 27 nak\d{s}atras.
Such a coordinate system implies a 
calculation where whole cycles are subtracted
from large numbers. Such modular arithmetic
appears to lie at the basis of the idea
of a superconjunction.
Traditionally, the 
Ved\={a}\.{n}ga Jyoti\d{s}a has been dated
to around 1350 BC, but a new paper
by Narahari Achar$^9$ argues for a much earlier
date of 1800 BC.

Van der Waerden$^{10}$
has argued that a primitive epicycle theory
was known to the Greeks by the time of Plato.
He argued such a theory might have been known in the
wider Indo-European world by early first millennium BC.
With new ideas about the pre-history of the
Indo-European world emerging, it is possible to
push this to an earlier millennium.
An old theory may be the source
which led to the development of very different epicycle
models in Greece and India.

The existence of an independent tradition of observation of
planets and a theory thereof as suggested by our analysis of the
\'{S}atapatha Br\={a}hma\d{n}a helps explain the puzzle why the classical
Indian astronomy of the Siddh\={a}nta period uses many constants that
are different from those of the Greeks.

\section{More on the Great Year}

Since the yuga in the Vedic and the Br\={a}hma\d{n}a periods
is so clearly obtained from an attempt to
harmonize the solar and the lunar years, it
appears that the consideration of the
periods of the planets was the basis of the creation of an
even longer yuga.

There is no reason to assume that the periods
of the five planets were unknown during
the Br\={a}hma\d{n}a age. I have
argued that the astronomical numbers in
the organization of the \d{R}gveda 
indicate with high probability the knowledge
of these periods in the \d{R}gvedic era
itself.$^{11}$ 

Given these periods, and the various yugas related
to the reconciliation of the lunar and the solar years,
we can see how the
least common multiple of these periods will
define a still larger yuga.

The Mah\={a}bh\={a}rata and the Pur\={a}\d{n}as
speak of the kalpa, the day of Brahm\={a},
which is 4,320 million years long.
The night is of equal length, and 360
such days and nights constitute a ``year'' of
Brahm\={a}, and his life is 100 such years
long.
The largest cycle is 311,040,000 million years
long at the end of which the world is
absorbed within Brahman, until another cycle
of creation. 
A return to the initial conditions (implying
a superconjunction) is inherent in such
a conception.
Since the Indians and the Persians were
in continuing cultural contact, it 
is certain that this old tradition 
became a part of the heritage of the Persians.
This explains how we come across the idea of the World-Year
of 360,000 years in the work of Ab\={u} Ma'shar, who also
mentioned a planetary conjunction in February 3102 BC. 

The theory of the transmission of
the Great Year of 432,000 years,
devised by Berossos, a priest in
a Babylonian temple, to India in about 300 BC,
was advanced by Pingree.$^{12}$
But we see this number being used in
relation to the Great Year in the
\'{S}atapatha Br\={a}hma\d{n}a itself, a long time
before Berossos.$^{13}$

The idea of superconjunction seems to be at the basis
of the cyclic calendar systems in India.
The  \'{S}atapatha Br\={a}hma\d{n}a speaks of a marriage between the
Seven Sages, the stars of the Ursa Major, and the K\d{r}ttik\={a}s;
this is elaborated in the Pur\={a}\d{n}as where it is stated that the
\d{r}\d{s}is remain for a hundred years in each nak\d{s}atra.
In other words, during the earliest times in India there existed a
centennial calendar with a cycle of 2,700 years.
Called the Saptar\d{s}i calendar,
it is still in use in several parts of India. Its current beginning is taken
to
be 3076 BE.

The usage of this calendar more than 2000 years ago is 
confirmed by the
notices of the Greek historians Pliny and
Arrian who suggest that, during the
Mauryan times, the Indian calendar began in 6676 BC.
It seems quite certain that this was the Saptar\d{s}i calendar with
a beginning which starts 3600 years earlier than
the current Saptar\d{s}i calendar.

The existence of a real cyclic calendar shows that the idea
of superconjunction was a part of the Indic tradition
much before the time of Berossos.
This idea was used elsewhere as well but, given the
paucity of sources, it is not possible to trace a
definite place of origin for it.
\section{Conclusions}

More than thirty years ago, Roger Billard showed$^{14}$ the falsity of 
the 19th century notion that India did not
have observational astronomy.
His analysis of the Siddh\={a}ntic and the practical kara\d{n}a
texts 
demonstrated that these texts
provide a set of elements from which the planetary
positions for future times can be computed.
The first step in these computations is the determination of the
mean longitudes which are assumed to be linear functions of time.
Three more functions, the vernal equinox, the lunar node and the
lunar apogee are also defined.

Billard investigated these linear functions for the five planets,
two for the sun (including the vernal equinox) and three for the
moon.
He checked these calculations against the values derived from
modern theory and he found that the texts provide very accurate
values for the epochs when they were written.
Since the Siddh\={a}nta and the kara\d{n}a models are not accurate,
beyond these epochs deviations build up.
In other words, Billard refuted the theory that there was no
tradition of observational astronomy in India.
But Billard's book is not easily available in
India, which is why 
the earlier theory has continued to do rounds
in Indian literature.

\={A}ryabha\d{t}a's constants are more accurate than
the one's available in the West at that time. He
took old Indic notions of the Great Yuga and of
cyclic time (implying superconjunction) and 
created a very original and novel siddh\={a}nta.
He presented  the rotation information with respect
to the sun which means that his system
was heliocentric to a certain extent.$^{15}$
Furthermore, he considered the earth to be
rotating on its own axis.
Since we don't see such an advanced system amongst
the Babylonians prior to the time of \={A}ryabha\d{t}a,
it is not reasonable to look outside of
the Indic tradition or \={A}ryabha\d{t}a himself
for the 
data on which these ideas were based. 

\section*{Notes and References}

\begin{enumerate}

\item 
K.D. Abhyankar, ``Babylonian source of \={A}ryabha\d{t}a's
planetary constants.''
{\it Indian Journal of History of Science}
35: 185-188, 2000.

\item S. Kak, {\it The Astronomical Code of the \d{R}gveda.}
New Delhi: Munshiram Manoharlal, 2000, page 87.

\item K.S. Shukla and K.V. Sarma, {\em \={A}ryabha\d{t}\={\i}ya of
\={A}ryabha\d{t}a.}
New Delhi: Indian National Science Academy, 1976.

\item D. Pingree, ``Astronomy and astrology in India and Iran.''
{\it Isis} 54: 229-246, 1963.

\item B.L. van der Waerden, ``The great year in Greek, Persian
and Hindu astronomy.'' {\it Archive for History of Exact 
Sciences} 18: 359-384, 1978.

\item S. Kak, {\it The Astronomical Code of the \d{R}gveda.}
New Delhi: Munshiram Manoharlal, 2000;\\
S. Kak, ``Birth and early development of Indian
astronomy.'' In {\it Astronomy Across Cultures: The History
of Non-Western Astronomy}, H. Selin (ed.), pp. 303-340.
Dordrecht: Kluwer Academic Publishers, 2000.

\item S. Kak, ``The sun's orbit in the Br\={a}hma\d{n}as.''
{\it Indian Journal of History of Science}
33: 175-191, 1998.

\item S. Kak, ``The astronomy of the age of geometric altars.''
{\it Quarterly Journal of the Royal Astronomical Society} 36: 385-396, 1995.

\item B.N. Narahari Achar, ``A case for revising
the date of Ved\={a}\.{n}ga Jyoti\d{s}a.''
{\it Indian Journal of History of Science} 35: 173-183, 2000.

\item  B.L. van der Waerden, ``The earliest form of the epicycle
theory.'' {\it Journal for the History of Astronomy} 5:
175-185, 1974.

\item S. Kak, {\it The Astronomical Code of the \d{R}gveda.}
New Delhi: Munshiram Manoharlal, 2000.

\item See note 4, above.

\item B.N. Narahari Achar, ``On the astronomical
basis of the data of \'{S}atapatha Br\={a}hma\d{n}a:
a re-examination of Dikshit's theory.''
{\it Indian Journal of History of Science} 35: 1-19, 2000.


\item Roger Billard, {\it L'astronomie Indienne.}
Paris: Publications de l'ecole francaise d'extreme-orient, 1971;\\
B.L. van der Waerden, ``Two treatises on Indian
astronomy.'' {\it Journal for History of Astronomy}
11: 50-58, 1980.

\item H. Thurston, {\it Early Astronomy.}
New York: Springer-Verlag, 1994, page 188.

\end{enumerate}

\end{document}